\documentclass[11pt]{article}
\usepackage{arxiv}
\usepackage[utf8]{inputenc}
\usepackage[T1]{fontenc}

\usepackage[english]{babel}
\usepackage{graphicx}
\usepackage{subcaption}
\usepackage{tabularx}
\usepackage{enumitem}
\DeclareUnicodeCharacter{2212}{-}

\usepackage[dvipsnames]{xcolor}
\usepackage{tikz}
\usepackage{colortbl}

\usepackage{listings}

\usepackage{gensymb}
\usepackage{amssymb}
\usepackage{amsmath}
\usepackage{amsfonts}
\usepackage{bbm}
\usepackage{wasysym}
\usepackage{mathtools}
\usepackage{amsthm}

\usepackage{natbib}
\let\cite\citep

\usepackage{xpatch}
\xpretocmd{\eqref}{Eq.~}{}{}

\definecolor{codegreen}{rgb}{0,0.6,0}
\definecolor{codegray}{rgb}{0.5,0.5,0.5}
\definecolor{codepurple}{rgb}{0.58,0,0.82}
\definecolor{backcolour}{rgb}{0.95,0.95,0.92}

\lstdefinestyle{mystyle}{
    backgroundcolor=\color{backcolour},   
    commentstyle=\color{codegreen},
    keywordstyle=\color{magenta},
    numberstyle=\tiny\color{codegray},
    stringstyle=\color{codepurple},
    basicstyle=\ttfamily\footnotesize,
    breakatwhitespace=false,         
    breaklines=true,                 
    captionpos=b,                    
    keepspaces=true,                 
    numbers=left,                    
    numbersep=5pt,                  
    showspaces=false,                
    showstringspaces=false,
    showtabs=false,                  
    tabsize=2
}

\newtheorem{remark}{Remark}

\DeclareFontFamily{U}{mathx}{\hyphenchar\font45}
\DeclareFontShape{U}{mathx}{m}{n}{<-> mathx10}{}
\DeclareSymbolFont{mathx}{U}{mathx}{m}{n}
\DeclareMathAccent{\widebar}{0}{mathx}{"73}

\DeclareUnicodeCharacter{0301}{\'{e}}
\DeclareUnicodeCharacter{1EF3}{\'{y}}

\usepackage{hyperref} 
\hypersetup{
    linktoc=all,
    colorlinks=true,
    linkcolor=black,
    citecolor=black,
    urlcolor=black,
}

\title{Assessing Time Series Correlation Significance: A Parametric Approach with Application to Physiological Signals}

\author{\href{https://orcid.org/0000-0002-7558-2071}{\includegraphics[scale=0.06]{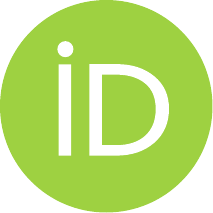}\hspace{1mm}Johan Medrano}      \vspace{2mm}  \\
	The Wellcome Centre for Human Neuroimaging \\
	UCL Queen Square Institute of Neurology \\
        London, WC1N 3AR, UK \vspace{2mm}  \\ 
	\texttt{\href{email:johan.medrano@ucl.ac.uk}{johan.medrano@ucl.ac.uk}} \\
        \And 
        \href{https://orcid.org/0000-0001-9033-9742}{\includegraphics[scale=0.06]{orcid.pdf}\hspace{1mm}Abderrahmane Kheddar} \vspace{2mm}  \\
	CNRS - AIST Joint Robotics Laboratory, IRL3218 \\
	Tsukuba 305-8560, Japan  \vspace{2mm}  \\
	CNRS - University of Montpellier LIRMM, UMR5506 \\
	Interactive Digital Human \\
        Montpellier, France \vspace{2mm}  
        \And 
        \href{https://orcid.org/0000-0001-8925-3546}{\includegraphics[scale=0.06]{orcid.pdf}\hspace{1mm}Sofiane Ramdani} \vspace{2mm} \\
	CNRS - University of Montpellier LIRMM, UMR5506 \\
	Interactive Digital Human \\
        Montpellier, France 
 }

\date{}

\begin{document}

\maketitle

\begin{abstract}
    {Correlation coefficients play a pivotal role in quantifying linear relationships between random variables. Yet, their application to time series data is very challenging due to temporal dependencies. This paper introduces a novel approach to estimate the statistical significance of correlation coefficients in time series data, addressing the limitations of traditional methods based on the concept of effective degrees of freedom (or effective sample size, ESS). These effective degrees of freedom represent the independent sample size that would yield comparable test statistics under the assumption of no temporal correlation. We propose to assume a parametric Gaussian form for the autocorrelation function. We show that this assumption, motivated by a Laplace approximation, enables a simple estimator of the ESS that depends only on the temporal derivatives of the time series. Through numerical experiments, we show that the proposed approach yields accurate statistics while significantly reducing computational overhead.  In addition, we evaluate the adequacy of our approach on real physiological signals, for assessing the connectivity measures in electrophysiology and detecting correlated arm movements in motion capture data. Our methodology provides a simple tool for researchers working with time series data, enabling robust hypothesis testing in the presence of temporal dependencies.}
\end{abstract}


\section{Introduction}
    \label{sec:intro}
    Correlation coefficients, such as Pearson's or Spearman's, are fundamental tools for assessing linear relationships between random variables. Although originally not designed for dependent samples, these coefficients are widely used with time series in diverse fields. Notable examples can be found in neuroimaging, where Pearson's correlation coefficient is used to construct functional brain networks from functional Magnetic Resonance Imaging (fMRI) data \cite{fox2005human,van2010intrinsic}, electrophysiology data \cite{zhang2021lightweight,naira2019classification,ji2019correlation,zhong2022spatio} or to relate brain signals to other behavioral parameters, such as movement data \cite{lu2021evaluation}. In this paper, we address the estimation of the statistical significance of correlation coefficients for time series.
    
    Fisher's variance-stabilizing transformation is useful to obtain simple bounds and significance of a correlation coefficient $r$. Under the null hypothesis that two sets of $n$ independent data points are uncorrelated, the null distribution of the Fisher-transformed variable $z=\text{arctanh}(r)$ is approximately normally distributed with mean $0$ and standard deviation $1/\sqrt{n-3}$. In other words, we have
    \begin{align}
        \label{eq:fisherdistr}
        \sqrt{n-3} \; \text{arctanh}(r) \sim \mathcal{N}\left(0, 1\right)
    \end{align}
    
    In real time-series data, the assumption of pairwise independence underlying traditional correlation coefficient tests often falters due to dependencies between consecutive observations. When data exhibit temporal dependencies, \eqref{eq:fisherdistr} is overconfident and fails to account for the loss of degrees of freedom due to temporal correlations. A pivotal solution, pioneered by Bartlett in 1935 \cite{bartlett1935some}, adjusts for this change of degrees of freedom by introducing a number of ``effective degrees of freedom'', or ``effective sample size'' (ESS), representing the independent sample size producing comparable test statistics. Under the null hypothesis that two sets of $n$ dependent data points, with ESS $\nu$ ($\nu\leq n$), are uncorrelated,  we can simply rewrite \eqref{eq:fisherdistr} to obtain the approximate distribution as:
    \begin{align}
        \label{eq:fisherdistredf}
        \sqrt{\nu-3} \; \text{arctanh}(r) \sim \mathcal{N}\left(0, 1\right)
    \end{align}
    This enables hypothesis testing even when actual samples show temporal dependencies. 
    
    The approach pioneered by Bartlett, that had various extensions, e.g.,~\cite{bartlett1946theoretical,bayley1946effective,quenouille1947notes,pyper1998comparison,afyouni2019effective}, relies on estimating the ESS from the sum of the product of the autocorrelation functions (ACFs) of the time series. Let $\rho_{k}$ and $\gamma_{k}$ be the autocorrelation at lag $k$ of two time series, resp.\ $x$ and $y$, of $n$ samples each. The ESS of the correlation coefficient between $x$ and $y$ is \cite{quenouille1947notes}:
    \begin{align}
        \label{eq:ess}
        \nu = n \left(\rho_0 \gamma_0  + 2\sum_{k=1}^{n-1} \frac{(n-k)}{n} \rho_{k} \gamma_{k} \right)^{-1} 
    \end{align}
    While this estimator of the ESS has been widely adopted, it requires computing the sample ACF which can be computationally demanding and can also yield inaccurate estimates of the ESS due to accumulating noise (in the sum of~\eqref{eq:ess}). 
    
    In this work, we assume a parametric form for the ACF to simplify the computation and estimation of the ESS. In particular, we show that the sum in \eqref{eq:ess} converges to an integral which can be analytically solved under a Gaussian (Laplace) approximation. The resulting expression relates the ESS to the average second spectral moment, also called roughness, of the pair of time series to correlate. Importantly, the roughness of a series can be estimated from the variance of its temporal derivatives. This scaffolds the central result of this work: for two processes $x$ and $y$ of length $n$ with temporal derivatives $\dot{x}$ and $\dot{y}$, the number of effective degrees of freedoms $\nu$ is approximately:
    \begin{align}
    \label{eq:edf-asymp}
    \nu = n \sqrt{\frac{\text{var}{(\dot{x})} + \text{var}{(\dot{y})}}{2\pi}}
    \end{align}
    We show that this formula yields accurate estimation of the ESS and effectively provides a computationally effective method to evaluate the significance of correlation coefficients.
    
    This article is organized as follows. First, we present the key contributions of our work, which encompass the derivation of the asymptotic expression for ESS, a straightforward approximation for Gaussian autocorrelation, and the introduction of an estimator based on the variance of the temporal derivatives of the processes. Then, we validate our approach through numerical experiments, demonstrating its effectiveness and its computational efficacy. Finally, we apply our methodology to real data to (i) highlight the significance of power-based connectivity measures in electrophysiology, and (ii) distinguish correlated and uncorrelated arm captured motion data.

\section{A parametric estimator of the effective sample size}
    Our work builds on estimating the ESS using a Gaussian approximation to the ACF. In this section we show that the ESS converges to an integral; then that a Laplace approximation of this integral relates the ESS to the signals roughness; and finally, that common roughness estimators can be used to build an estimator of the ESS.

    This work focuses on estimating the ESS for the correlation of smooth, wide-sense stationary (WSS) stochastic processes. A smooth WSS stochastic process $x$ is defined by convolving  white noise $w$ with a smooth ($C^\infty$), square-integrable function $K:\mathbb{R}\to\mathbb{R}^+$, in other word: 
    \begin{align}
        x(t) = \int K(t-s) w(s) ds 
    \end{align}
    
    For clarity and brevity, we reuse the same notation throughout this paper. By default $x$ and $y$ are two smooth WSS stochastic processes with zero mean, variances resp.\ $\sigma_x$ and $\sigma_y$, and autocorrelations functions $\rho$ and $\gamma$. For simplicity, we restrict our result section to the case where both $x$ and $y$ have the same autocorrelation, i.e.,\ $\rho = \gamma$. This restriction is moreover motivated in Section~\ref{sec:gaussacf}.

    \subsection{Asymptotic expression for the ESS}
        \label{sec:asymp}
        This section introduces an asymptotic form for the ESS. We consider~\eqref{eq:ess} for the ESS in the case of infinitely large $n$  and infinitely small sampling interval of the ACF. Under these conditions, the sum in \eqref{eq:ess} converges to 
        \begin{align}
        \label{eq:integral}
        \lim_{n\to+\infty}\;\; \frac{n}{\nu} = \int_{-\infty}^{+\infty} \rho(\tau) \gamma(\tau) d\tau 
        \end{align}
        
        This yields a new asymptotic ESS expression, $\nu_\infty$: 
        \begin{align}
            \label{eq:asymp}
            \nu_{\infty} = n \left(\int_{-\infty}^{+\infty} \rho(\tau) \gamma(\tau) d\tau \right)^{-1}
        \end{align}    
        
        The resulting integral formulation is pivotal in deriving analytical expressions for the ESS of stochastic processes with known ACF. As for processes with known autocorrelation function, we can directly evaluate~\eqref{eq:asymp}.  In the next section, we develop a generic approximation based on a Laplace approximation of the integral. 

    \subsection{Closed-form expression under a Laplace approximation}
        \label{sec:gaussacf}
        A general approach to approximate the integral in the asymptotic ESS expression is to use Laplace's method. In this section, we derive a general approximation to asymptotic ESS (\eqref{eq:asymp}), based on a Laplace approximation of the integral~(\eqref{eq:integral}). In brief, Laplace's method is used to approximate the integral of a arbitrary function with a unique global maximum using a simpler, Gaussian function around its peak \cite{penny2011statistical}. Typically in physiological signals, ACFs have a mode at lag~$0$ and decrease to $0$ for large lags, although they might exhibit long-range correlations. Thus, the area under the curve of the product of two such ACFs is expected to be contained around $0$, with an attenuation of long-range correlations. This justifies using Laplace's method to approximate the asymptotic ESS expression. 
                        
To derive the expression under Laplace approximation, note that the second-order expansion of the ACFs $\rho$ and $\gamma$ around their mode at $\tau = 0$ is similar to processes with Gaussian ACFs
        \begin{align}
            \hat{\rho}(\tau) = \exp({-\frac{1}{2}|\rho''(0)| \tau^2}) &&
            \hat{\gamma}(\tau) = \exp({-\frac{1}{2}|\gamma''(0)| \tau^2})
        \end{align}
        where $\rho''(0)$ and $\gamma''(0)$ are the second spectral moment of the processes, i.e., the second-order derivatives of the ACFs at $0$.   The second spectral moment is a universal measure of ``roughness'' in the literature of stochastic processes \cite{friston2008variational,cox2017theory}. From this Gaussian form, we can explicitly evaluate the integral in \eqref{eq:asymp}: 
        \begin{align}
            \label{eq:denominator-ext-ws}
            \int_{-\infty}^{\infty} \hat{\rho}(\tau) \hat{\gamma}(\tau)  d\tau =  \sqrt{\frac{2\pi}{|\rho''(0)| + |\gamma''(0)|}}
        \end{align}  By substituting~\eqref{eq:denominator-ext-ws} in~\eqref{eq:asymp}, the ESS under a Laplace approximation is  
        \begin{align}
            \label{eq:ext-ws-laplace}
            \nu_{\infty} = n \sqrt{\frac{|\rho''(0)|+|\gamma''(0)|}{2\pi}}
        \end{align}
        To summarise, we can use a Laplace (Gaussian) approximation to approximate the ESS. This is equivalent to fitting the ACF using a one-parameter family of Gaussian functions parameterised by their second spectral moment, or roughness. This allows to derive a simple expression for the ESS that only involves the second spectral moment of each process.  

        \begin{remark}
            \label{remark:acf}
            When $\rho''(0) = \gamma''(0)$, \eqref{eq:ext-ws-laplace} reduces to $n \sqrt{|\rho''(0)|/\pi}$.
            When $\rho''(0) \neq \gamma''(0)$, the ESS of the pair of signals given by~\eqref{eq:ext-ws-laplace} is similar to that of a fictive pair of signals with equal roughness $(|\rho''(0)| + |\gamma''(0)|)/2$. Thus, there is no need to consider furthermore the case with heterogeneous roughness, as it can be mathematically reduced to the case with a (homogeneous) roughness given by the arithmetic mean of the original roughnesses.  
        \end{remark}
        
    \subsection{Estimating the second spectral moment}
         A well-established result from stochastic process theory is that the roughness can be conveniently estimated from the variance of the first-order temporal derivatives of the process \cite{cox2017theory, adler2010geometry, worsley1996unbiased}. In this section, we leverage this result to construct an estimator of the asymptotic ESS.  
         
         Let $x$ and $y$ be two processes with autocorrelation functions $\rho$ and $\gamma$ and temporal derivatives $\dot{x}$ and $\dot{y}$. The second spectral moments $|\rho''(0)|$ and $|\gamma''(0)|$ can be estimated from the variance of the temporal derivatives of the process, i.e., $|\rho''(0)| = \text{var}(\dot{x})$ and $|\gamma''(0)| = \text{var}(\dot{y})$. Pluging in \eqref{eq:ext-ws-laplace}, we obtain an estimator of the ESS:  
         \begin{align}
                \label{eq:ess-gaussian}
             \nu_{\infty} = n \sqrt{\frac{\text{var}(\dot{x}) + \text{var}(\dot{y})}{2 \pi}}
         \end{align}
         By construction, this new parametric estimator for the ESS depends only on the variance of the temporal derivatives of the time series.  As such, its consistency and unbiasedness derive directly from those of the variance estimators (from direct application of the continuous mapping theorem). 
         
         \begin{remark}
             The second spectral moment is also related to the expected number of zero-crossings of a process by Rice's formula~\cite{rice1944mathematical}. For didactic purposes, we show an alternative ESS estimator based on Rice's formula in Appendix~\ref{sec:ess-rice}.
         \end{remark}

\section{Numerical validation}
    We conduct numerical experiments on Gaussian processes with Gaussian autocorrelation functions (GPGA) to assess the validity and limitations of our approach. The numerical results are organised in four parts. First, we evaluate the quality of fitting the squared autocorrelations from the process roughness. Second, we compare the quality of the estimates of the ESS, corresponding to the integral of the squared ACFs, with existing methods. Then, we compare the statistics yielded by our method with existing ones. Finally, we investigate the computational performance of our approach.   
    
    \subsection{Assessment of parametric autocorrelation function estimates}
    The method we present leverages the analytical evaluation of the ESS of a pair of stochastic processes from their roughness, estimated from the variance of the temporal derivatives. Here, we verify whether this method can accurately estimate the ACF of the process, when the ACF is known to be Gaussian.
    
    We generate 2000 sample paths from GPGA with different roughness levels. We then retrieve the squared autocorrelation using the variance of the derivatives. To sample a sample path from a GPGA with roughness $r$, we convolve a unit white Gaussian noise sample with a Gaussian kernel of variance $2/r$ as per \cite{friston2008variational}. We then apply \eqref{eq:edf-asymp} to determine its second spectral moment from the variance of the process temporal derivatives.
    
    \begin{figure}[t]
        \centering
            \begin{subfigure}{\textwidth}
            \begin{subfigure}{0.32\textwidth}
                \includegraphics[width=\linewidth]{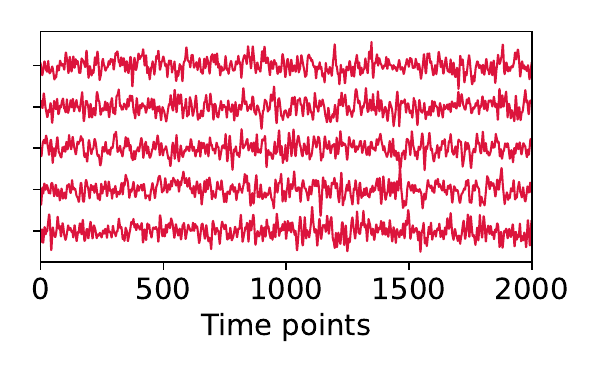}
                \caption{Sample paths, $\rho''(0) = 10^{-1}$}
            \end{subfigure}
            \begin{subfigure}{0.32\textwidth}
                \includegraphics[width=\linewidth]{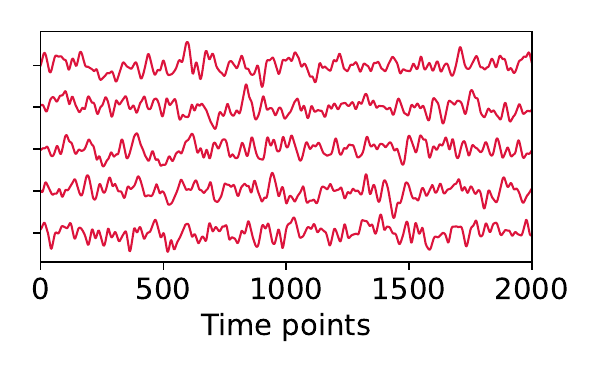}
                \caption{Sample paths, $\rho''(0) = 10^{-2}$}
            \end{subfigure}
            \begin{subfigure}{0.32\textwidth}
                \includegraphics[width=\linewidth]{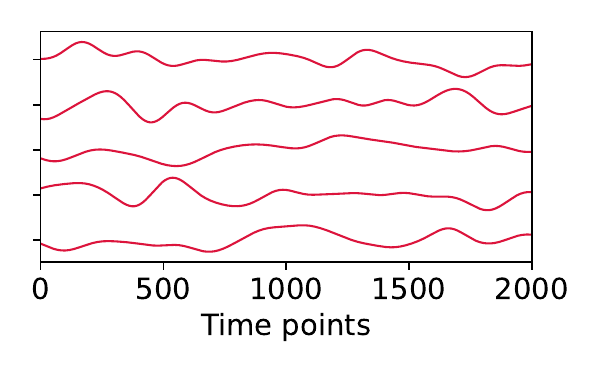}
                \caption{Sample paths, $\rho''(0) = 10^{-4}$}
            \end{subfigure}
        \end{subfigure}
        \begin{subfigure}{\textwidth}
            \begin{subfigure}{0.32\textwidth}
                \includegraphics[width=\linewidth]{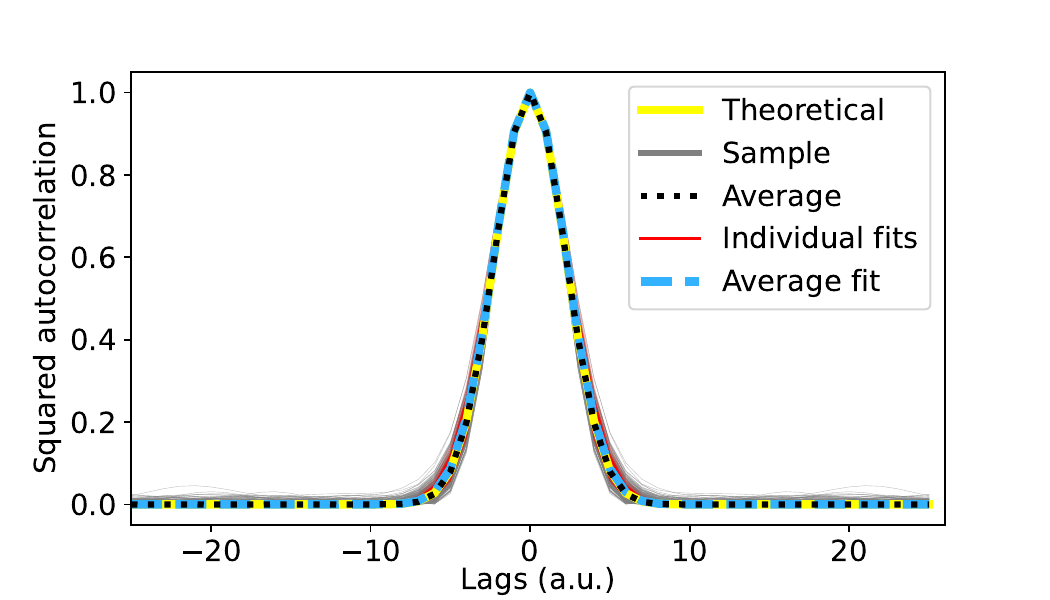}
                \caption{Squared ACF, $\rho''(0) = 10^{-1}$}
            \end{subfigure}
            \begin{subfigure}{0.32\textwidth}
                \includegraphics[width=\linewidth]{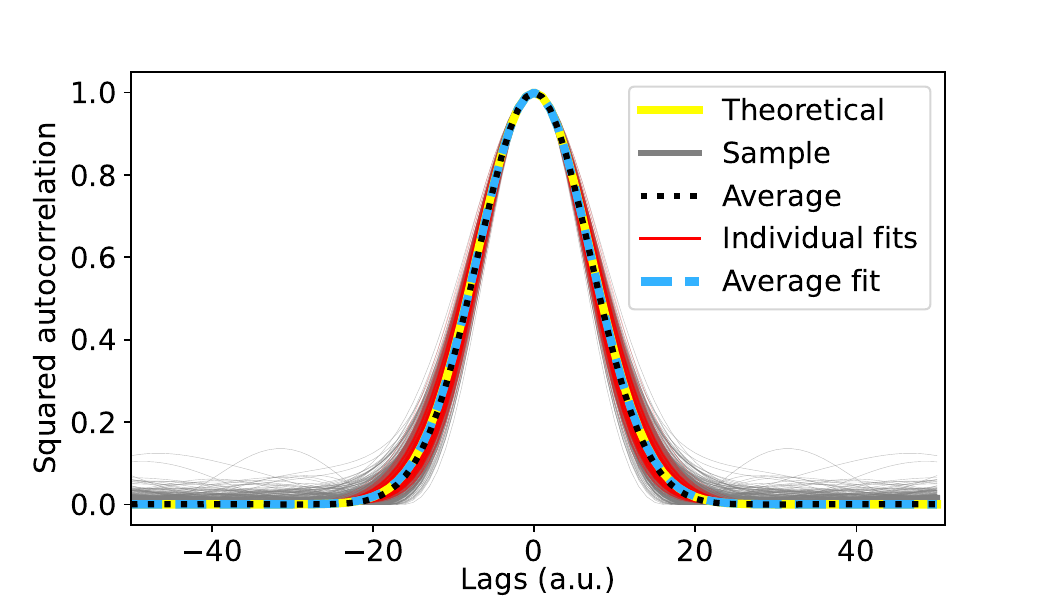}
                \caption{Squared ACF, $\rho''(0) = 10^{-2}$}
            \end{subfigure}
            \begin{subfigure}{0.32\textwidth}
                \includegraphics[width=\linewidth]{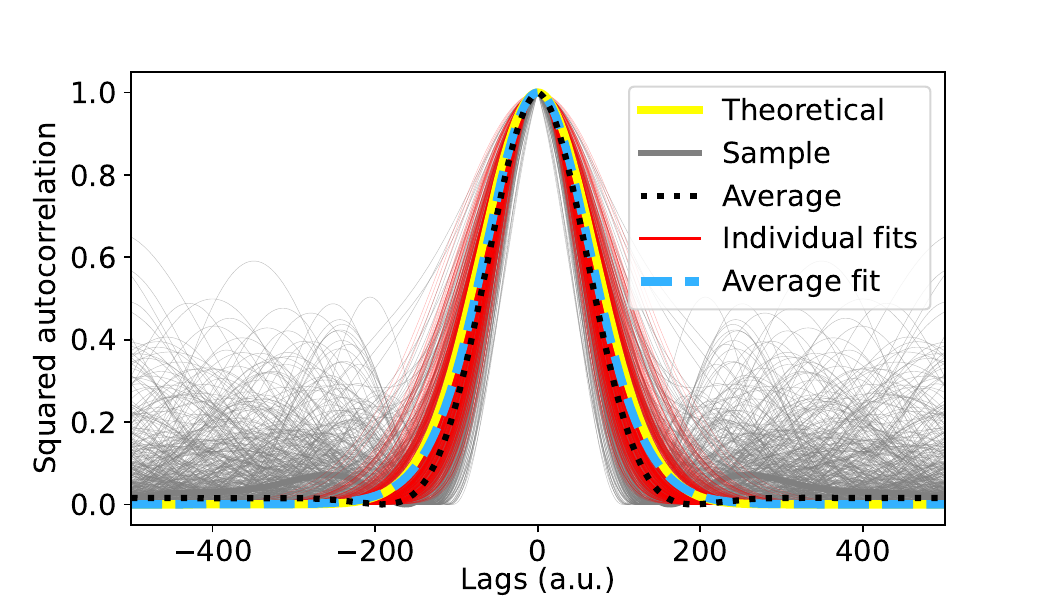}
                \caption{Squared ACF, $\rho''(0) = 10^{-4}$}
            \end{subfigure}
        \end{subfigure}
        \caption{Sample paths and squared autocorrelation functions for roughness values $\rho''(0)$. Rows one and two display, respectively, samples of Gaussian processes for different roughness values and their squared autocorrelation functions. Notably, the squared ACF plays a crucial role in ESS estimation. The plots showcase various ACF estimations: theoretical (plain yellow), sample (plain grey), average of sample (dotted black), Gaussian fit based on estimated roughness (plain red), and Gaussian fit based on average roughness (dashed blue). The bias in sample ACFs, evident in the grey curves, increases as roughness decreases due to increasing random long-range correlations. On the other hand, the Gaussian fits (plain red) remain unbiased at large lags, ensuring the ACF retains its Gaussian form. }
        
        \label{fig:samples-acf}
    \end{figure}
    
    Figure \ref{fig:samples-acf} illustrates that low roughness results in significant biases in squared sample autocorrelation functions, leading to ESS underestimation. This bias becomes clear in Figure~\ref{fig:samples-acf}(f), where the grey curves diverge significantly from their expected near-zero values. This figure underscores the merit of predetermining the functional form of the ACF, thereby filtering out spurious long-term correlations. Furthermore, our method consistently aligns more closely with the theoretical ACF relatively to the direct sample squared ACF.
    
    In conclusion, explicitly formulating the functional form of the ACF filters out errant long-term correlations otherwise amplified by traditional ESS estimators based on sample autocorrelations. This leads to more accurate and consistent ACF estimates across varying roughness scales. However, we note that the variance of the squared ACF appears to be inversely proportional to roughness. This issue is further investigated in subsequent sections.

        \subsection{Assessment of parametric ESS estimates}
            This section deals with the influence of the roughness estimator's variance on the variance of the estimated sum of squared autocorrelation and that of the ESS. We also examine how these relate to the length of series.
            
\begin{figure}[t]
                \centering
                \begin{subfigure}{\textwidth}
                    \begin{subfigure}{0.24\textwidth}
                        \includegraphics[width=\linewidth]{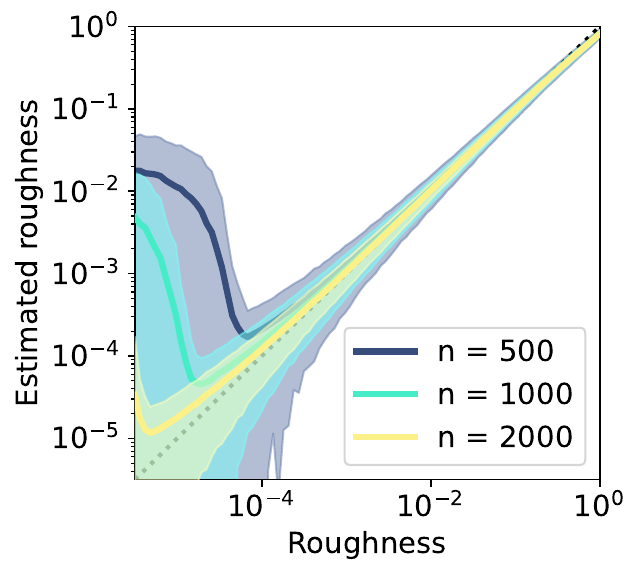}
                        \caption{Roughness}
                    \end{subfigure}
                    \begin{subfigure}{0.24\textwidth}
                        \includegraphics[width=\linewidth]{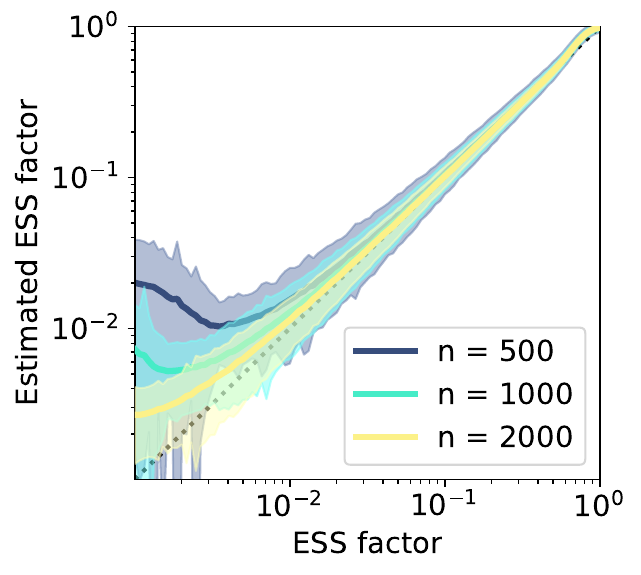}
                        \caption{ESS - FFT-based}
                    \end{subfigure}
                    \begin{subfigure}{0.24\textwidth}
                        \includegraphics[width=\linewidth]{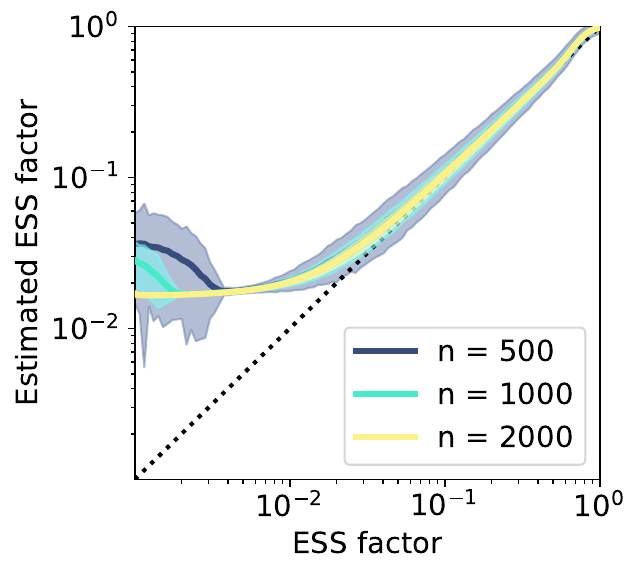}
                        \caption{ESS - Welch-based}
                    \end{subfigure}
                    \begin{subfigure}{0.24\textwidth}
                        \includegraphics[width=\linewidth]{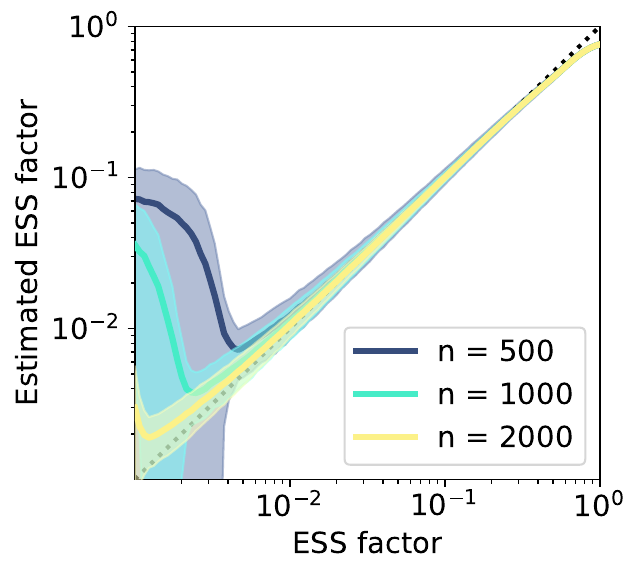}
                        \caption{ESS - Proposed}
                    \end{subfigure}
                \end{subfigure}
                \caption{a) Estimated roughness as a function of the process roughness, for series of 500 points (dark blue), 1000 points (light blue), and 2000 points (yellow). b-d) Estimated ESS factor as a function of the process true ESS, computed from \eqref{eq:ext-ws-laplace}, for series of 500 points (dark blue), 1000 points (light blue), and 2000 points (yellow) for the FFT-based (b), Welch-based (c), and proposed approaches (d). Shaded areas correspond to the $95\%$ confidence intervals. }
                \label{fig:ess}
            \end{figure}

            We sampled 1000 sample paths from GPGA, adjusting roughness between $10^{-6}$ and $1$. Each sample path's roughness is then estimated. Using this roughness estimate, we compute the ESS using~\eqref{eq:ess-gaussian}. Furthermore, we determine the ACF of the process using the inverse Fourier transform of the sample path's power spectral density (PSD). The latter is estimated either through the FFT or the Welch periodogram with a 256-point window having 128-point overlap~\cite{welch1967use}. The ESS is then obtained from the ACF using~\eqref{eq:ess}. We replicate this procedure for series lengths of $500$, $1000$, and $2000$ points. For comparison, the ESS is normalized by series length, yielding an ESS factor. Figure~\ref{fig:ess} contrasts estimated and theoretical roughness, as well as the ESS factors from various methods against those from theoretical roughness.
            
            From Figure~\ref{fig:ess}(a), it is clear that roughness estimates share a similar trend. For minuscule roughness values, bias emerges. Extended series better approximate low roughness, hinting that biases result from series that are too short relative to their roughness. As expected from their relationship (\eqref{eq:asymp}), ESS factors derived from roughness mirror this trend.
            
            Interestingly, ESS factors from both FFT and Welch-based squared ACF methods also exhibit similar biases. The Welch periodogram's bias appears even at elevated ESS factors due to its fixed window size. From a statistical standpoint, biases and high variances are expected when approaching a near-zero sample size.
            
            Conclusively, our method offers a closer ESS estimate to true values than traditional methods, especially for low ESS factors corresponding to low roughness. This confirms the ability of our method to derive the ESS from process temporal derivative variances, showing at superior correlation coefficient statistic accuracy, which we confirm in the subsequent section.

    \subsection{Statistics under the null hypothesis}
        We have previously demonstrated the efficiency of the proposed method in fitting the squared autocorrelation function of a process and its superior accuracy in ESS estimation across varied roughness scales. Now, we assess how this improved ESS estimation influences the computation of probabilities under the null hypothesis, i.e., the $p$-value, especially under two distinct roughness scales.
        
\begin{figure}[t]
            \centering
            
            \begin{subfigure}{\textwidth}
                \begin{subfigure}{0.32\textwidth}
                    \includegraphics[width=\linewidth]{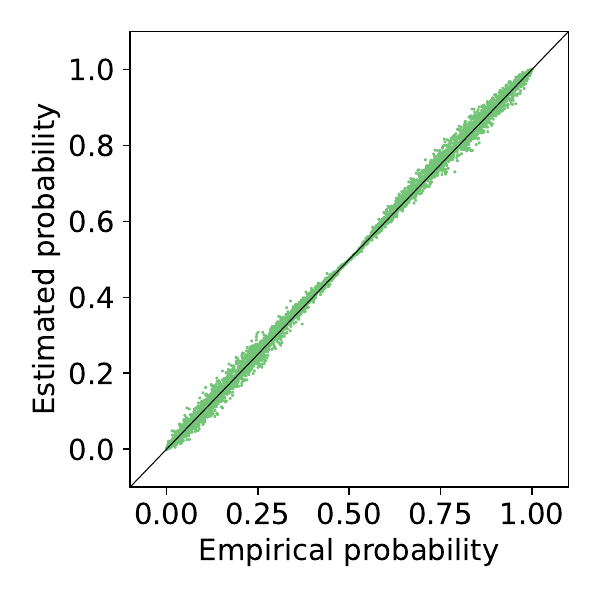}
                    \caption{FFT-based, $\rho''(0) = 10^{-2}$}
                \end{subfigure}
                \begin{subfigure}{0.32\textwidth}
                    \includegraphics[width=\linewidth]{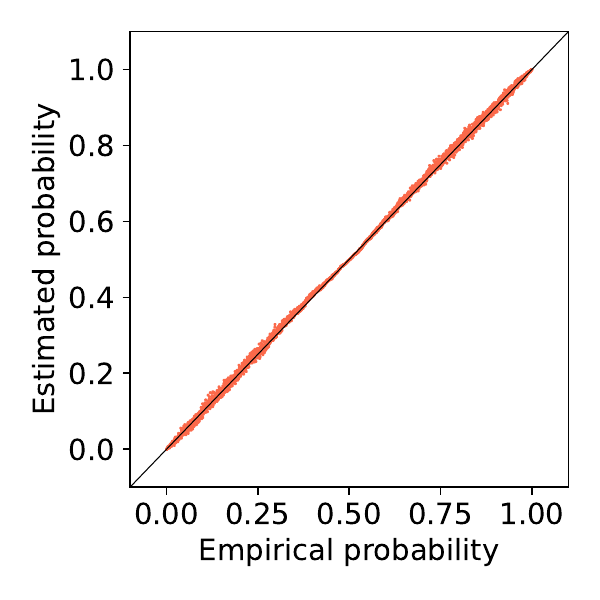}
                    \caption{Welch-based, $\rho''(0) = 10^{-2}$}
                \end{subfigure}
                \begin{subfigure}{0.32\textwidth}
                    \includegraphics[width=\linewidth]{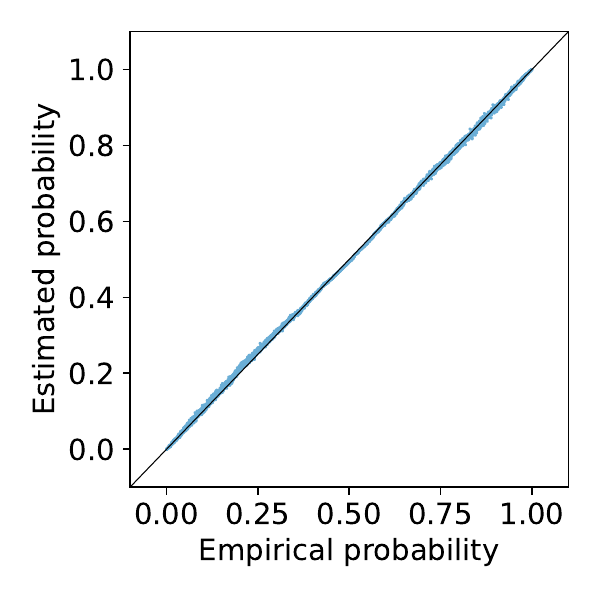}
                    \caption{Proposed, $\rho''(0) = 10^{-2}$}
                \end{subfigure}
            \end{subfigure}
            \begin{subfigure}{\textwidth}
                \begin{subfigure}{0.32\textwidth}
                    \includegraphics[width=\linewidth]{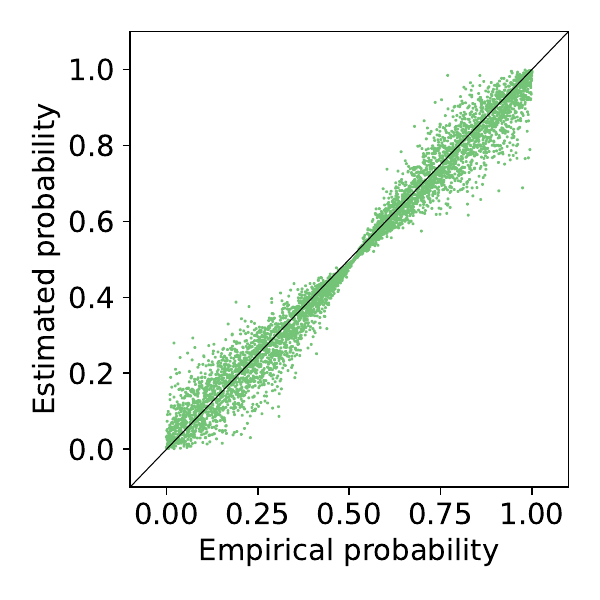}
                    \caption{FFT-based, $\rho''(0) = 10^{-4}$}
                \end{subfigure}
                \begin{subfigure}{0.32\textwidth}
                    \includegraphics[width=\linewidth]{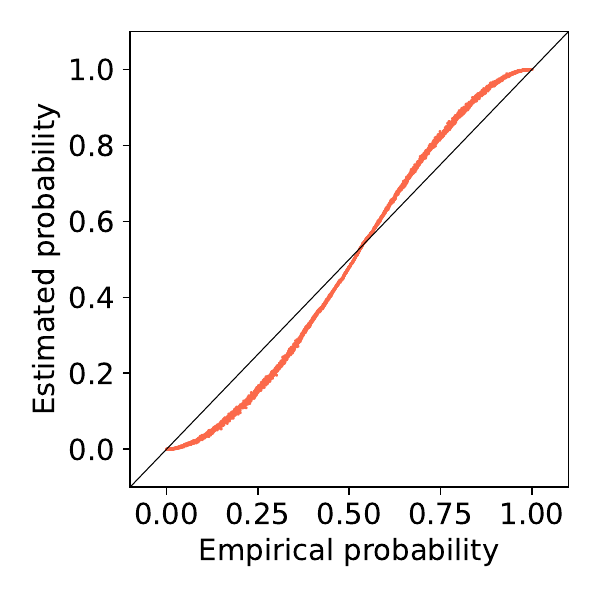}
                    \caption{Welch-based, $\rho''(0) = 10^{-4}$}
                \end{subfigure}
                \begin{subfigure}{0.32\textwidth}
                    \includegraphics[width=\linewidth]{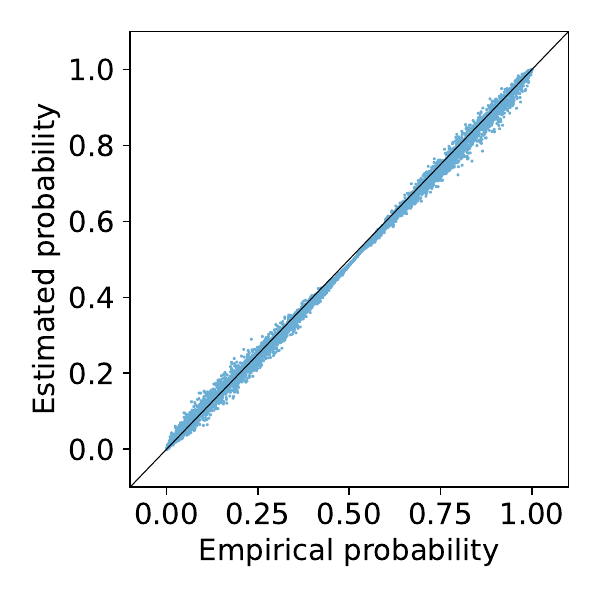}
                    \caption{Proposed, $\rho''(0) = 10^{-4}$}
                \end{subfigure}
            \end{subfigure}
            \caption{Estimated probability against empirical probability for the different approaches. The first row illustrates a roughness of $10^{-2}$, while the second showcases $10^{-4}$. Each column represents different estimation methods: FFT-based (green), Welch-based (red), and our method (blue).}
            \label{fig:pp-plot-numerical}
        \end{figure}

        We set our method against those reliant on the sum of squared autocorrelation, where ACF is computed via FFT or the Welch periodogram. We generated 5000 pairs of 2000-points-long sample paths from processes with specified roughness and determined their correlation coefficient. Being inherently uncorrelated, these sample paths effectively sample the null distribution of GPGA at that roughness. Consequently, the empirical probability of each sample correlation coefficient converges to its probability under the null hypothesis. Each sample's empirical probability can be compared with the probability yielded by any of the methods. These results are represented as probability-probability plots in Figure~\ref{fig:pp-plot-numerical} for roughness values of $10^{-2}$ and $10^{-4}$.
        
        At a roughness of $10^{-2}$, the methods show comparable statistical outputs. The FFT-based $p$-values display more dispersion than both proposed and Welch-based methods, which are markedly analogous. Reducing roughness to $10^{-4}$ modifies the statistics for each method. Both our method and the FFT-based exhibit increased $p$-value variability. In contrast, the Welch method displays pronounced bias with minimal variability.
        
        The trends in Figure~\ref{fig:pp-plot-numerical} are more comprehensible when compared with the ESS estimates presented in  Figure~\ref{fig:ess}. The transition from a roughness of $10^{-2}$ to $10^{-4}$ equates to a shift from an ESS factor of roughly $10^{-1}$ to $10^{-2}$. Such changes mirror the variances and biases seen in our ESS estimates, which directly affect $p$-value computation. Given that our method consistently provides superior ESS estimations across roughness values, it is hence inferred to offer enhanced statistics.
        
        In summary, our numerical analyses confirm the paramountcy of a parametric autocorrelation function, where the unique parameter is estimated via the process's second spectral moment. This method circumvents biases introduced by random long-range dependencies, resulting in more accurate sample $p$-values for correlation coefficients. In subsequent sections, we advocate for GPGA's aptness in real-world applications, such as neuroimaging data connectivity analysis and movement trajectory correlations.

    \subsection{Comparison of computational performance}
        In this section, we evaluate the performance gain of our approach. The main difference with existing methods is that we avoid explicit evaluation of the ACF by using a parametric ACF, which only requires to estimate the average series roughness. 

        To evaluate the impact of series length on the computation time and speedup, we generate pairs of sample paths from a GPGA process with roughness $10^{-3}$  having varying length, from 100 to $10^{7}$ points with 10 fold increments. For each pairs of paths, we compute the ESS using FFT-based and Welch-based computation of the ACFs, and with our approach. The FFT-based ACF is obtained using \texttt{numpy}'s \texttt{fft} and \texttt{ifft} functions \cite{harris2020array}. The Welch-based ACF is computed using \texttt{scipy}'s \texttt{welch} function from the \texttt{signal} package with a window length of 256 points (100 points for the 100 points series)  and \texttt{numpy}'s \texttt{ifft} \cite{2020SciPy-NMeth}. Our  approach uses \texttt{numpy}'s \texttt{diff} and \texttt{var} to compute the variance of the temporal derivatives of the process. The results are obtained on a laptop equipped with an 12-core Intel Xeon W-10855M CPU and 32Gb of RAM running Ubuntu~22.04 and Python~3.8. For each approach, we computed the timings from 100 loops and present the mean and standard deviation across 7 different runs.

    \begin{figure}[t]
        \centering
            \begin{subfigure}{\textwidth}
            \begin{subfigure}{0.49\textwidth}
                \includegraphics[width=\linewidth]{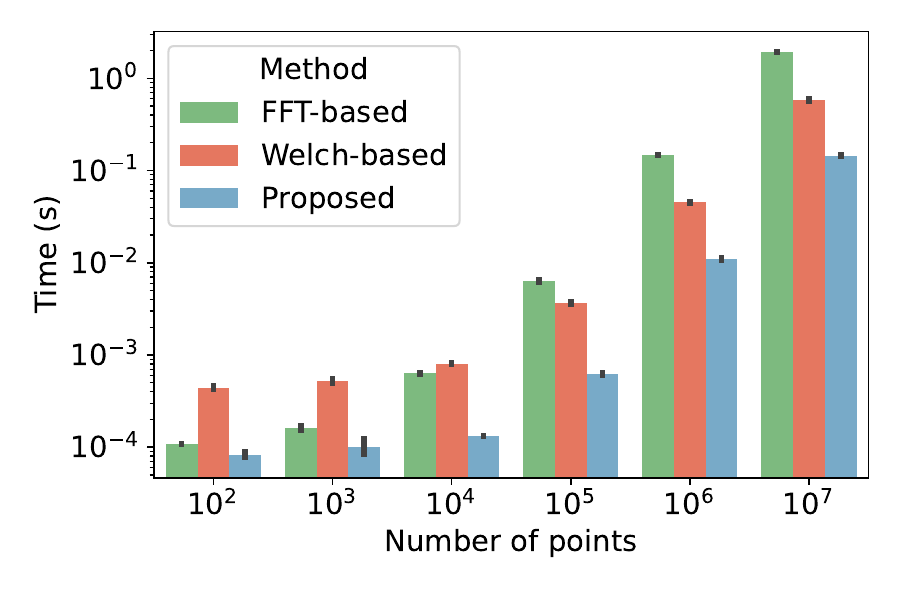}
                \caption{Computation times}
            \end{subfigure}
            \begin{subfigure}{0.49\textwidth}
                \includegraphics[width=\linewidth]{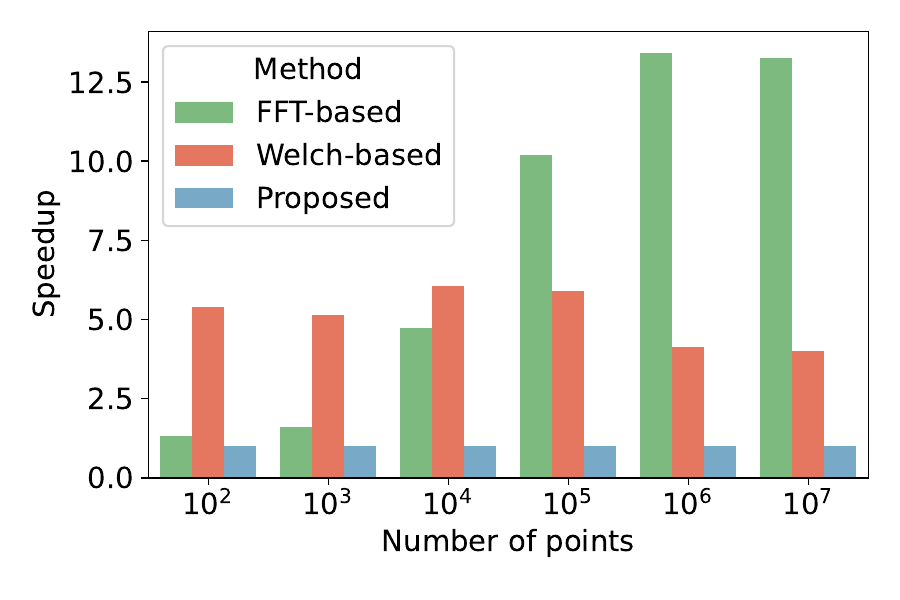}
                \caption{Speedup of proposed approach}
            \end{subfigure}
        \end{subfigure}
        \caption{a) Comparison of the computation time for varying number of points from 100 to $10^7$. b) Speedup of our approach, obtain as the average computation time of each method divided by the average computation time of the proposed method. For instance, a speedup of 5 means that our approach divides by 5 the average computation time.}
        
        \label{fig:computation} 
    \end{figure}

        We report the results on Figure~\ref{fig:computation}. The average computation times obtained by the proposed approach are significantly below that of FFT-based and Welch-based methods. This result is consistent across all different number of points. Looking at the speedup confirms the computational advantage of the proposed approach.  
        In average across different number of points, the proposed approach is 7.4 times faster than an FFT-based approach. This speedup is negligible for short time series (resp. 1.3 and 1.6 for 100 and 1000 points) but significant for long time series (10.2 to 13.4 for $10^5$ to $10^7$). The speedup from Welch-based approach is relatively constant across time series length due to the fixed window length, with an average speedup of 5.1.
        
        To sum-up, our proposed approach, based on the temporal derivatives of the time series, gives a significant speedup as compared to other approaches based on the computation of the ACF. This computational argument motivates using a parametric Gaussian form of the ACF in applications where timing is critical or where large amounts of data have to be processed; for instance, when evaluating correlations of brain activity between a large number of brain regions.

\section{Application to physiological signals}
    \subsection{Use-case~1: Evaluating power-based connectivity from electrophysiology data}
        \label{sec:exeeg}
        \subsubsection{Context and motivation}
        Electroencephalography (EEG) captures the electric potentials on the scalp. Its low-cost and high-temporal-resolution make EEG a particularly appealing  neuroimaging technique to evaluate the functional connectivity between brain regions. A standard approach, power-based connectivity analysis, measures functional connectivity by evaluating the correlation between the time-frequency power of the signal at two different locations \cite{cohen2014analyzing}. Several methods exist to perform power-based connectivity analysis, but the correlation between wavelet-based time-frequency representations is particularly interesting within the scope of this work. 
        
\begin{figure}
                \centering
                \includegraphics[width=\textwidth]{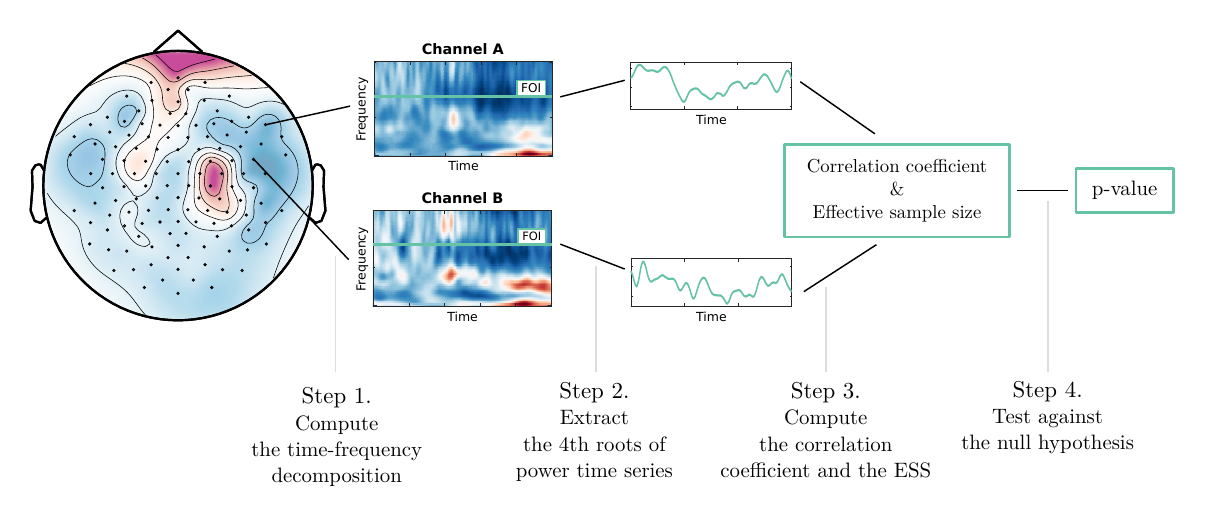}
                \caption{Procedure to assess the connectivity of two EEG channels at a particular frequency of interest (FOI).}
                \label{fig:eeg-procedure}
            \end{figure}

        The time course of power can be obtained using a continuous wavelet transform with a Morlet wavelet. A Morlet wavelet combines a Gaussian kernel with a complex sinusoid at a specified frequency. Therefore, a continuous wavelet transform with a Morlet wavelet introduces a form of temporal smoothing with a Gaussian kernel. Intuitively, if the width of the Gaussian smoothing kernel is large enough relative to the autocorrelation of the signal, the resulting time series will have roughly a Gaussian autocorrelation. In this case, one can consider testing the strength of the correlation between two power time series under the null hypothesis that the signals are two uncorrelated GPGA. The overall procedure is illustrated in Figure~\ref{fig:eeg-procedure}. This approach allows a more efficient analysis of connectivity significance without relying on computationally demanding methods. 

        This section aims at gaining some insight on whether the null hypothesis of uncorrelated GPGA processes might yield appropriate statistics. In contrast to the numerical experiments section, we cannot sample from the null distribution of correlation coefficient between band-power time series. Thus, we can only investigate a few characteristics of the data to assess the validity of our approach. Indeed, this section echoes the first step of statistical analysis in real world application, where statisticians have no access to the data generative distribution and shall assess the suitability of their model.    

    \subsubsection{Data presentation and preprocessing}
       We analyse a dataset containing EEG recordings of a subject listening to continuous, naturalistic speech \cite{di2015low, crosse2016multivariate}. The subject listened to a classic work read by an English speaker. EEG signals were recorded from a single participant with a Biosemi system having 128 channels and a sampling rate of 512~Hz. Prior to analysis, the data has been filtered between 1 and 15~Hz and downsampled to 128Hz. We then computed the time-frequency representation of the EEG signals using a continuous wavelet transform. We used Morlet wavelet with 7 cycles to produce time-frequency representation with a sufficient temporal smoothness to fall under our Gaussian assumptions. Finally, we take the fourth root of the power to make the data marginally Gaussian \cite{hawkins1986note}.

        \begin{remark}
            Note that selecting the number of cycles in the wavelet effectively selects the degree of smoothness of the signal. This shows that the number of cycles is related to the ESS. In Appendix~\ref{sec:ess-wavelet}, we give a formula giving the ESS as a function of the number of cycles.  
        \end{remark}

    \subsubsection{Adequacy of GPGA assumptions for wavelet-based EEG power}
        Here, we show that our approach yieds appropriate statistics for power-based connectivity analysis.  We first look at whether the signals have an approximately Gaussian autocorrelation and whether our approach provides a good fit of the squared autocorrelation function. We compare results given by FFT-based and Welch-based approaches with our approach, and analyse the results in the light of our numerical results.  Finally, we compare the ESS and 97.5\% quantile given by all three approaches with real data and sample GPGA sample paths with matching roughness.
        
        \begin{figure}[t!]
            \centering
                \begin{subfigure}[b]{0.32\textwidth}
                        \includegraphics[width=\linewidth, height=2.65in]{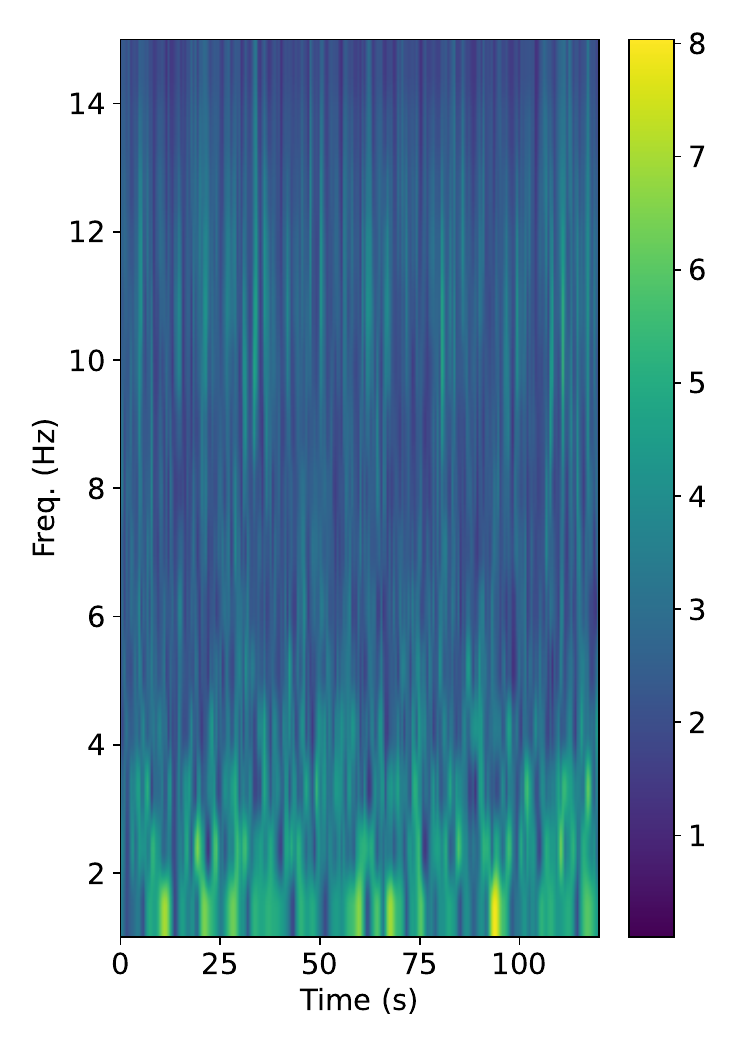}
                        \vspace{-2em}
                        \caption{Time-frequency plot}
                \end{subfigure}
                \vspace{1em}
                \begin{subfigure}[b]{0.65\textwidth}
                    \begin{subfigure}{\textwidth}
                        \includegraphics[width=\linewidth]{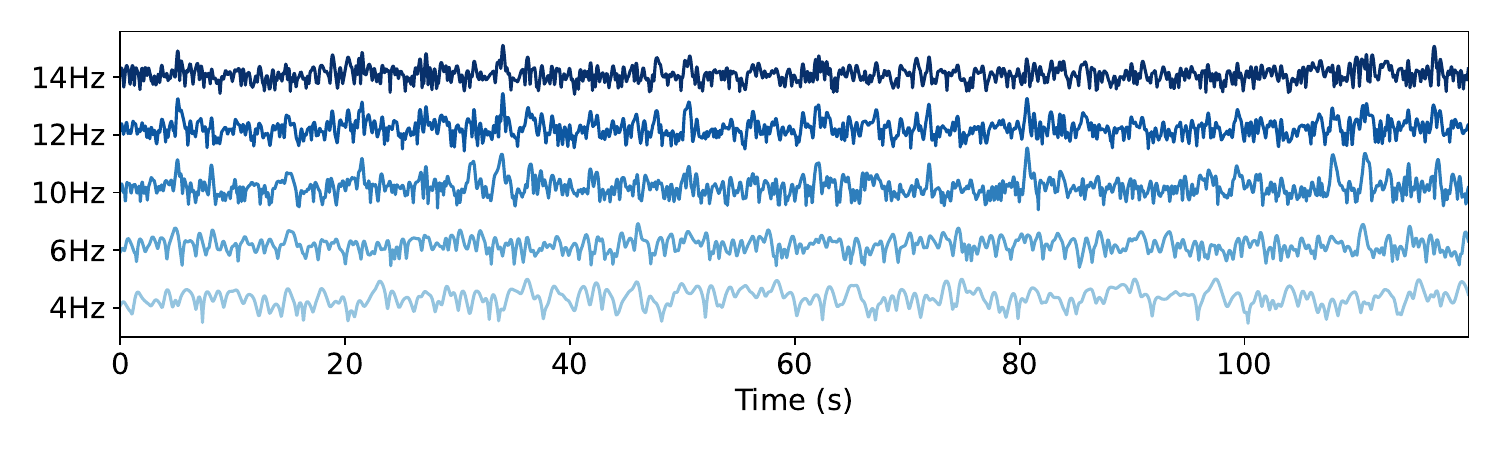}
                        \vspace{-2em}
                        \caption{Sample power time series}
                    \end{subfigure}
                    \begin{subfigure}{0.49\textwidth}
                        \includegraphics[width=\linewidth]{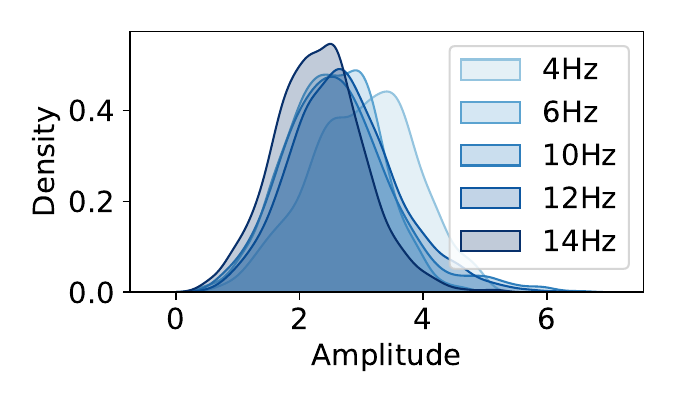}
                        \vspace{-2em}
                        \caption{Marginal distribution}
                    \end{subfigure}
                    \begin{subfigure}{0.49\textwidth}
                        \includegraphics[width=\linewidth]{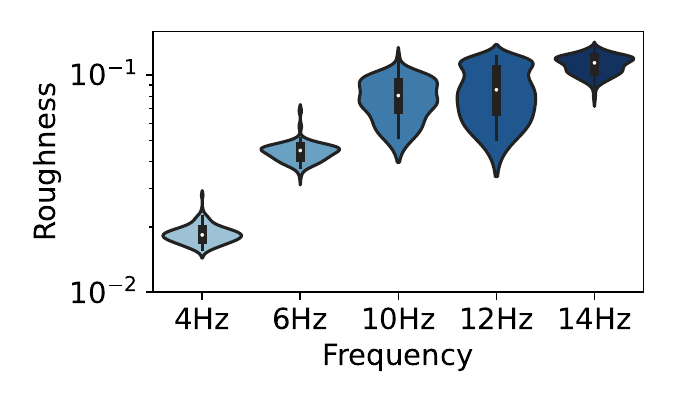}
                        \vspace{-2em}
                        \caption{Roughness}
                    \end{subfigure}
                \end{subfigure}
                \vspace{1em}
                \begin{subfigure}{\textwidth}
                    \centering
                    \begin{subfigure}{0.32\textwidth}
                        \includegraphics[width=\linewidth]{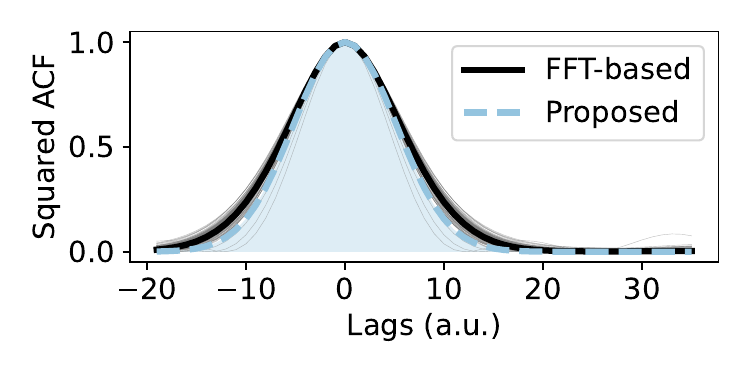}
                        \vspace{-2em}
                        \caption{Squared ACF, 4Hz}
                    \end{subfigure}
                    \begin{subfigure}{0.32\textwidth}
                        \includegraphics[width=\linewidth]{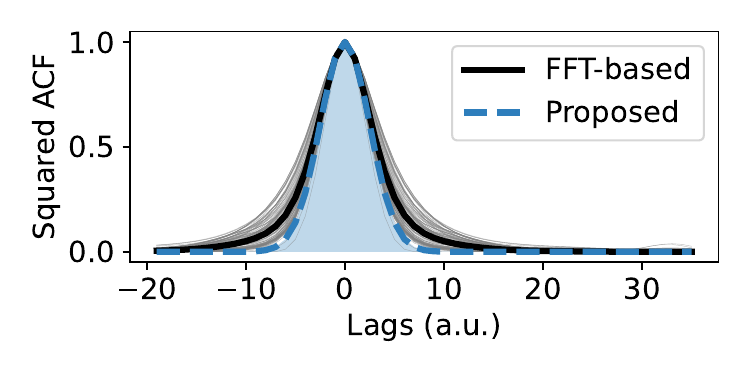}
                        \vspace{-2em}
                        \caption{Squared ACF, 10Hz}
                    \end{subfigure}
                    \begin{subfigure}{0.32\textwidth}
                        \includegraphics[width=\linewidth]{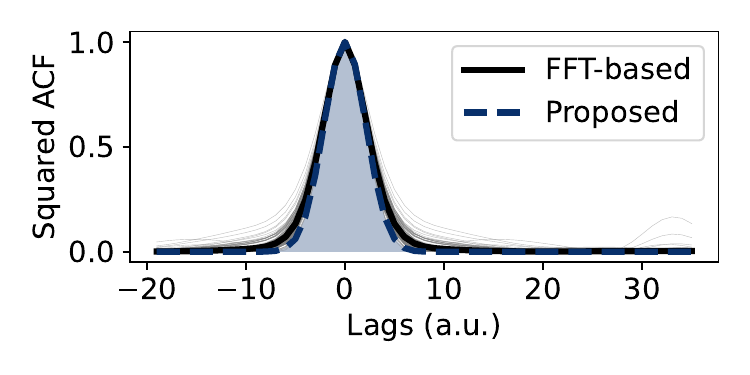}
                        \vspace{-2em}
                        \caption{Squared ACF, 14Hz}
                    \end{subfigure}
                \end{subfigure}
            \caption{(a) Time-frequency plot of the fourth root of power for channel A9. (b) Sample fourth-root of power time series at different frequencies. (c) Marginal distribution of fourth-root of power time series across channels. (d) Roughness across channels at different frequency. (e-g) Squared ACF at $4$~Hz, $10$~Hz, and $14$~Hz. Gray lines indicate different channels, the black line indicates the mean squared ACF across channels, the dotted blue line indicates the mean fit using the proposed approach, and the filled blue area shows the area that is integrated in the denominator of the ESS expression. }
            \label{fig:results-eeg}
        \end{figure}
        Figure~\ref{fig:results-eeg} summarizes the properties of the data. In particular, we show a time-frequency plot (Fig.~\ref{fig:results-eeg}~(a) and (b)) and some samples power time series at different frequencies. Intuitively, the time course of power at lower  frequencies is smoother than for higher frequencies. This is confirmed by looking at the roughness of the signals across channels, whose mean increase with frequency (Fig.~\ref{fig:results-eeg}~(d)). In addition, we display (Fig.~\ref{fig:results-eeg}~(c)) the  marginal distribution of the signals across channels to inspect its normality -- a key element in mandating using Pearson's correlation coefficient instead of Spearman's rank correlation. Finally, we inspect (Fig.~\ref{fig:results-eeg} (e), (f), and (g)) the squared ACF at $4$~Hz, $10$~Hz, and $14$~Hz, and compare it against the Gaussian fit obtained from the signals roughness. We observe that the squared ACF is roughly Gaussian, despite a slower decrease (lower kurtosis) which  might come from autocorrelations inherent to the nature of the signal. Overall, these results suggest that GPGA might provide a good approximation for the data and that our approach gives appropriate statistics.

    \subsubsection{Assessment of ESS and quantiles for wavelet-based EEG power}
\begin{figure}[!t]
            \centering
                \begin{subfigure}{\textwidth}
                \centering
                    \begin{subfigure}{0.45\textwidth}
                        \centering
                        \includegraphics[width=\linewidth]{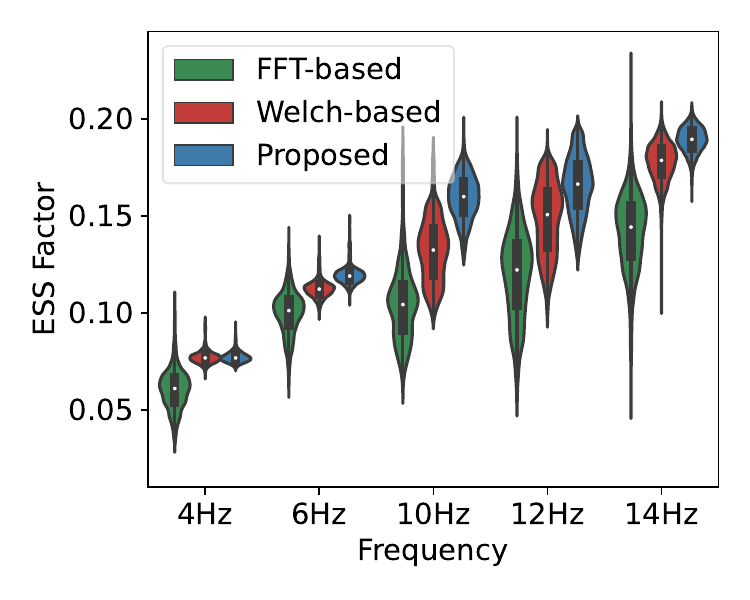}
                        \vspace{-2em}
                        \caption{ESS factor from power time series}
                    \end{subfigure}
                    \begin{subfigure}{0.45\textwidth}
                        \includegraphics[width=\linewidth]{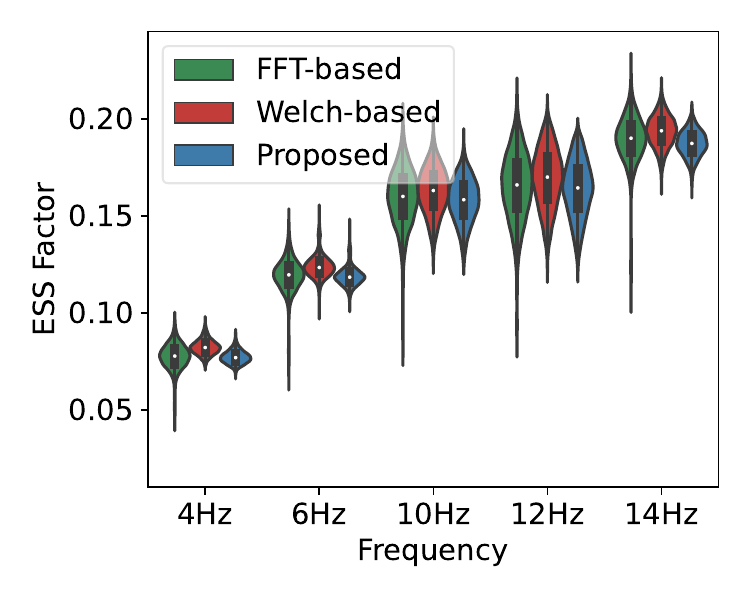}
                        \vspace{-2em}
                        \caption{ESS factor under GPGA assumptions}
                    \end{subfigure}
                \end{subfigure}
                \begin{subfigure}{\textwidth}
                \centering
                    \begin{subfigure}{0.45\textwidth}
                        \centering
                        \includegraphics[width=\linewidth]{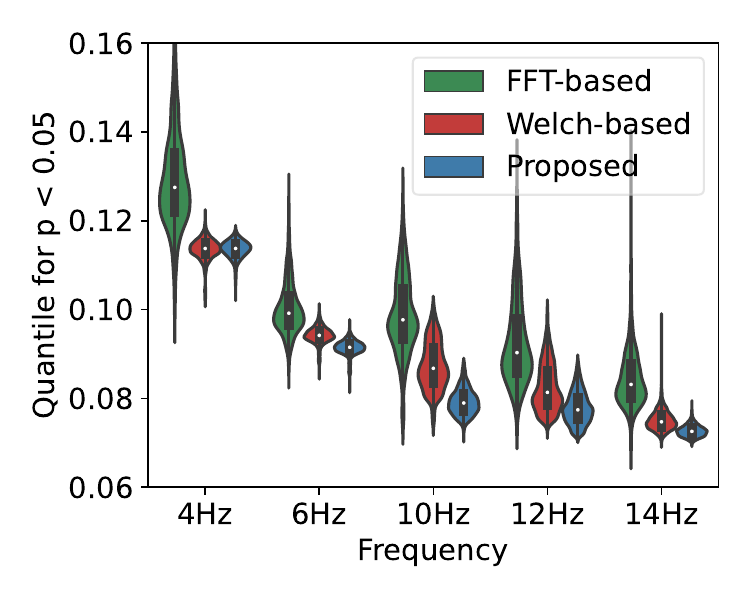}
                        \vspace{-2em}
                        \caption{97.5\% quantile for power time series}
                    \end{subfigure}
                    \begin{subfigure}{0.45\textwidth}
                        \includegraphics[width=\linewidth]{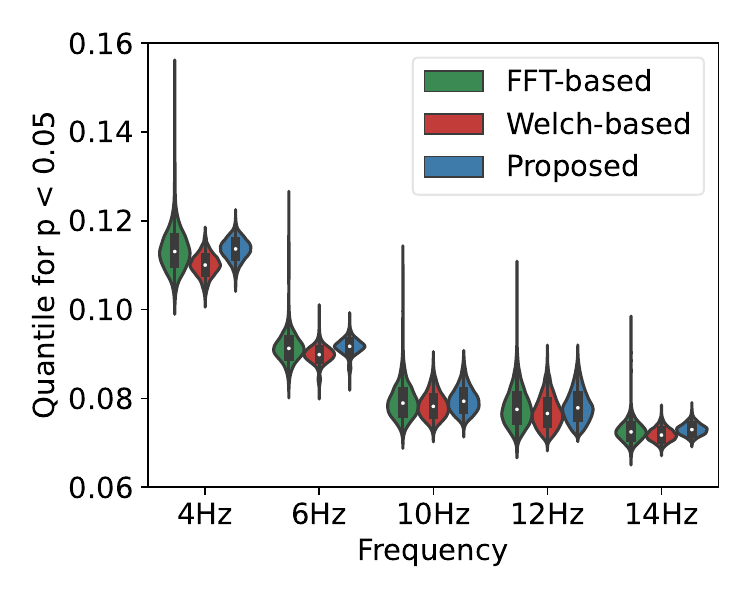}
                        \vspace{-2em}
                        \caption{97.5\% quantile under GPGA assumptions}
                    \end{subfigure}
                \end{subfigure}
            \caption{(a) Distribution of the ESS factor evaluated with the FFT-based (green), the Welch-based (red), and our approach (blue) for pair of power time series at a given frequency. (b) Similar to (a) for random GPGA sample paths with roughness matching that of every pair of signals at a given frequency. (c) Distribution of the $97.5$\% quantile obtained from the ESS of each pair of real time series.  (d) Similar to (c) for random GPGA sample paths with matching roughness.}
            \label{fig:results-eeg-2}
        \end{figure}

        To confirm these results, we evaluate the ESS factor for all possible between-channel pairs connectivity at each given frequency. In addition, for each signal we generate a random GPGA sample path with matching length and roughness, and compute the ESS factor for all pairs of random sample paths. This gives us a hint on how the ESS factor would behave if the sample paths where effectively sampled from the null hypothesis, i.e., GPGA with matching length and roughness. We complement this analysis by looking at the $97.5$\% quantile of both real and simulated data, under the null hypothesis. Any correlation coefficient greater than the $97.5$\% quantile -- or lower than its opposite -- would be effectively considered as significant with $p < 0.05$. The $97.5$\% quantile is computed by the inverse Fisher transform of the $97.5$\% quantile of the Gaussian distribution: 
        \begin{align}
        \label{eq:quantile}
            Q_{97.5\%} = \tanh(1.96/\sqrt{\nu_\infty})
        \end{align} The ESS factor and 97.5\% quantile for both real and simulated data are presented on Figure~\ref{fig:results-eeg-2}.  
        
        We observe that the ESS factor for real EEG signal increases with frequency. This is expected as the roughness of power time series increases with the frequency, as shown in Figure~\ref{fig:results-eeg}. This increased ESS factor results in a decreased $97.5$\% quantile. Indeed, as the effective sample size increases and statistical power increases, we can arbitrate on the rejection of the null hypothesis at lower values of correlation coefficient. This is conform to the role played by the ESS in~\eqref{eq:quantile}. 
        
        Interestingly, we see that both FFT-based and Welch-based approaches give lower values of ESS, and higher $97.5$\% quantile value, than our proposed approach. This effect is the strongest at $10$~Hz and $12$~Hz. Comparing each figure with its analogous generated under GPGA assumptions, we see that under GPGA assumptions all three methods yield approximately the same values of ESS factor and $97.5$\% quantile. Because these differences are stronger at $10$~Hz and $12$~Hz, i.e., in the well-identified alpha band, we hypothesis that task-related modulations of the power task cause variations in the autocorrelation of the power; which induce variability and shift in the ESS factor and related quantiles. Note that the size of this effect on the $97.5$\% quantile is relatively small (less than $0.02$). We argue that this is negligible in most cases, especially given the variability of the sample estimates of the roughness or ACFs.

    \subsection{Use-case 2: Detecting correlation between random movement trajectories}    
        This section investigates the suitability of our approach for biological signals. We analyse the statistics of movement trajectories as recorded by a motion capture system. The signals analysed in the present section are in nature complementary to the time course of EEG power analysed in the previous section. This is because they possess a very low roughness and are not marginally Gaussian. Thus, we aim at evaluating our proposed approach on less `well-behaved' biological signals. 

        The motion capture data analysed in this section comes from a single subject performing self-paced unilateral elbow flexion/extension. The data was acquired during a study on the EEG correlates of arm movements. The experimental procedure conformed the Declaration of Helsinki and was approved by the local ethics committee. The subject was instructed to move continuously in a self-paced pseudo-random manner for $40$ runs (one per side) of $23.5$~seconds. The position of body segments was recorded using an XSens Awinda suit at a $60$~Hz rate and the elbow flexion angle was obtained as the angle between the upper arm and forearm. 

\begin{figure}[htb!]
            \centering
                \begin{subfigure}[t]{\textwidth}
                    \begin{subfigure}[t]{0.32\textwidth}
                        \includegraphics[width=\linewidth]{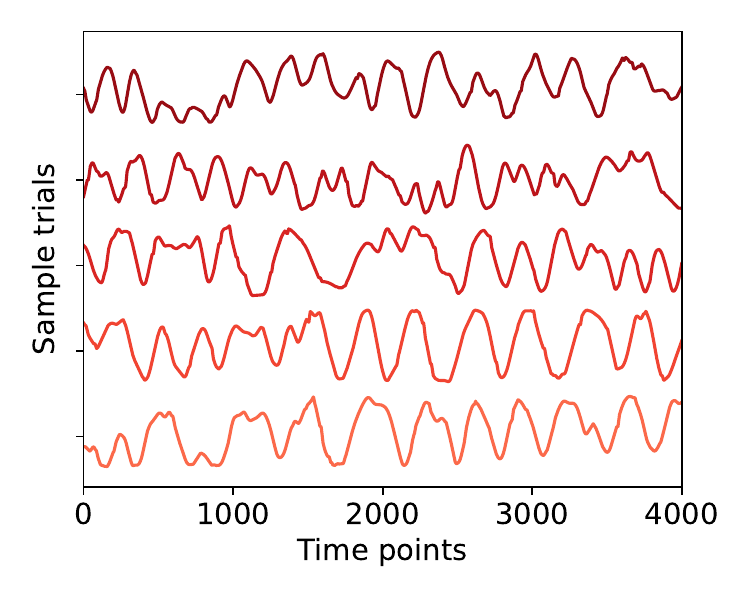}
                        \caption{Sample paths for various trials}
                    \end{subfigure}
                    \begin{subfigure}[t]{0.32\textwidth}
                        \includegraphics[width=\linewidth]{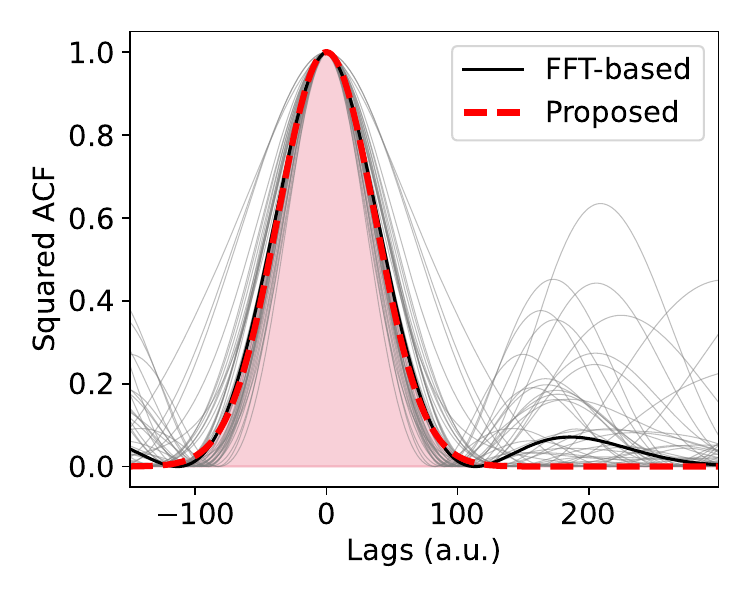}
                        \caption{Squared ACFs}
                    \end{subfigure}
                    \begin{subfigure}[t]{0.32\textwidth}                     \includegraphics[width=\linewidth]{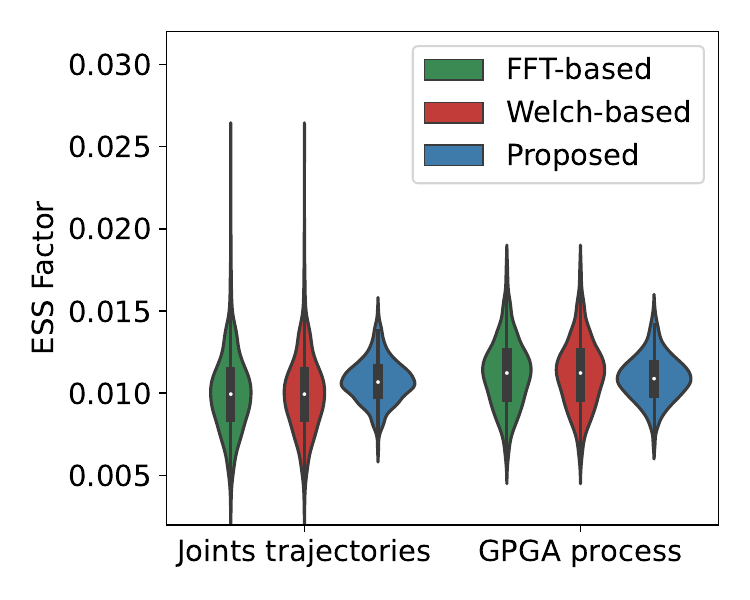}
                        \caption{ESS factors}
                    \end{subfigure}
                \end{subfigure}
                \vspace{1em}
                \begin{subfigure}{\textwidth}
                    \begin{subfigure}{0.32\textwidth}
                        \includegraphics[width=\linewidth]{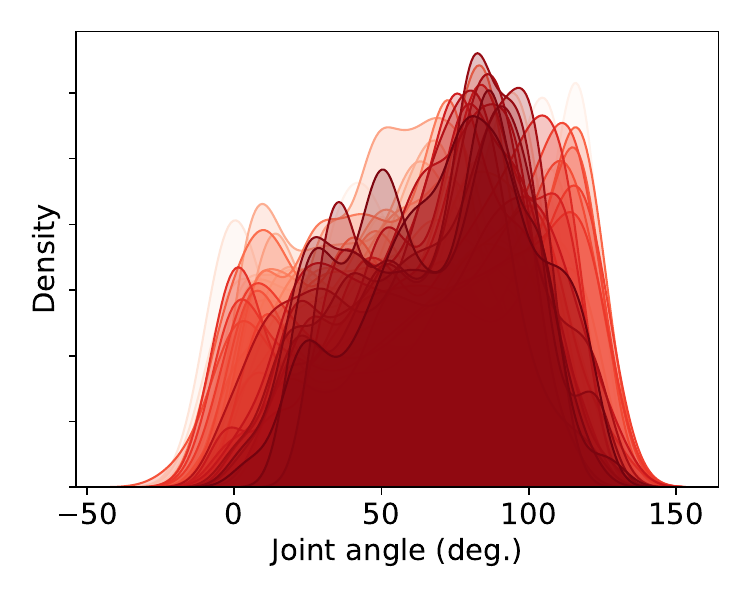}
                        \caption{Joint angle distribution}
                    \end{subfigure}
                    \begin{subfigure}{0.32\textwidth}
                        \includegraphics[width=\linewidth]{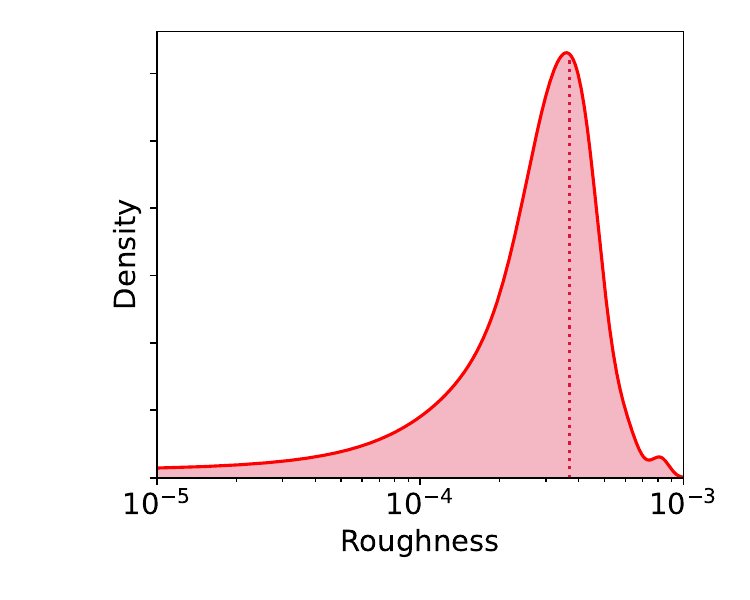}
                        \caption{Samples roughness}
                    \end{subfigure}
                    \begin{subfigure}{0.32\textwidth}
                        \includegraphics[width=\linewidth]{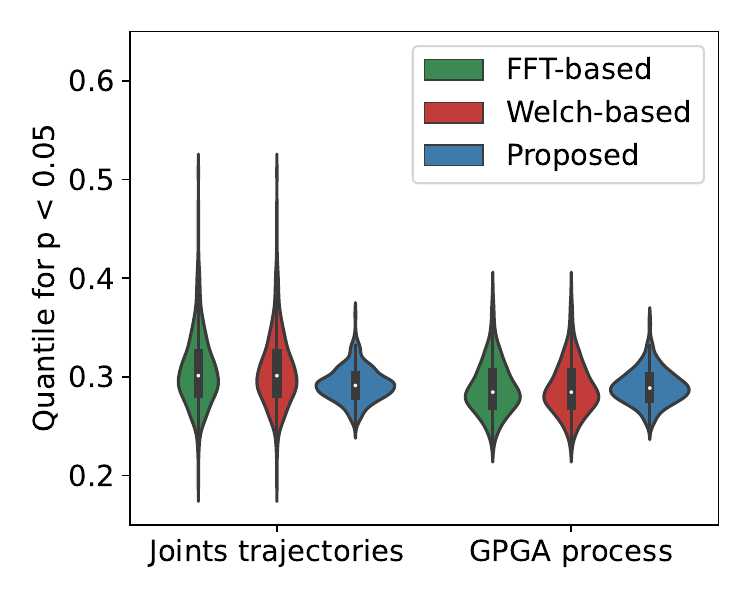}
                        \caption{$97.5$\% quantiles}
                    \end{subfigure}
                \end{subfigure}
            \caption{(a) Sample joint trajectories. (b) Squared ACFs computed from the FFT of the trials (gray), their average (black), and the Gaussian fit obtained from the average sample roughness (dashed red). The area under the curve (red) corresponds to the inverse of the mean ESS factor for the proposed approach. (c) ESS factors computed from joint trajectories and random GPGA sample paths with matching roughness, for all possible pairs of trials. (d) Marginal distribution of the joint angle for all trials. (e) Kernel density plot of the distribution of the roughness over trials. (f) $97.5$\% quantile computed from joint trajectories and random GPGA sample paths with matching roughness, for all possible pairs of trials.}
            \label{fig:results-joints}
        \end{figure}

        We computed the marginal distribution, squared ACF, and roughness of each trial. In addition, we computed the ESS factors and the $97.5$\% quantile for the correlation coefficient between every possible pair of trials. Similarly to the previous section, we generated for each trial a random GPGA sample path with matching roughness, and computed the same quantities as for the real data. Results are presented in Figure~\ref{fig:results-joints}. 

        Looking first at the statistics of the joint trajectories, we see that the joint angle distribution is not Gaussian, nor even unimodal. In practice, this would motivate to use Spearman's correlation coefficient instead of Pearson's. Additionally, we observe that the squared ACFs display  correlations for lags around $200$~points, that are not matching the Gaussian profile. On the other hand, the main peak of the squared ACF seem well approximated by the Gaussian fit. We also observe that the signal roughness is quite low, with an average roughness of $3.7\times 10^{-3}$. 

        Interestingly, the ESS factors and $97.5$\% quantiles computed for both the real joint trajectories and random GPGA processes are relatively similar, with differences being much smaller than the spread of the values. This result highlight that overall, the ESS of pairs of joint trajectories from this dataset are quite similar to that of GPGA with matching roughness. Consequently, quantiles are also relatively similar. This comparative results shows that, despite being not marginally Gaussian and having long-range autocorrelations not fitting to the Gaussian profile, the significance of the correlation between joint trajectories from this dataset could be accurately computed under GPGA assumptions.

\section{Conclusion and future directions}
    In this work, we considered the problem of testing for significant correlation between autocorrelated time series. Several works have proposed to use an ESS to correct the test distribution. The ESS correspond to the size of a sample of independent observations that would produce the same statistics. These approaches require to estimate the autocorrelation function, which can be problematic when testing correlation between two unique and smooth series.  

    Here, we derive the asymptotic expression of the ESS. This asymptotic expression can be used with parametric forms of ACF to analytically derive the ESS of a given process. In particular, we show that the expression for the ESS takes a simple form when considering a Gaussian approximation of the ACF. This simple expression depends on the second-order derivatives of the autocorrelation function at its mode, which is a measure of the roughness of the process and relates to several of its statistical properties. Roughness can be easily estimated from the variance of the first-order temporal derivatives of the process. 

    We conducted numerical experiments to validate the proposed method. We observe that our approach retrieves the statistics of correlation coefficients under the null hypothesis that signals are GPGA. For this particular type of process, our approach outperforms classical approaches, yielding robust estimate of the statistics under a large range of roughness. In addition, our approach shows relatively higher computational performances compared to alternative methods.  

    In a second part, we show that some data issued from biological signals satisfy GPGA assumptions. Our approach seem adequate to test for significant correlations between brain regions, when used with power-based connectivity analysis of electrophysiological data. Our method also applies well to random joint trajectories, as measured from a motion capture system. 

    Overall, obtained results suggest that our approach yields accurate and robust statistics under assumptions that can be found in real biological signal. More generally, we claim that using a parametric form of autocorrelation can yield more accurate statistics, without being overly restrictive on the form of the process. In particular, we think that more advanced methods for parameter estimation could allow better estimation of the autocorrelation function. However, as often with parametric approaches, there is no one-size-fits-all form of the autocorrelation function, and the statistician holds the responsibility to evaluate the adequacy of the model for the data under study. 

    In addition to correlation coefficients, our approach can be straightforwardly extended to linear regression. A pivotal assumption of linear regression is the independence of residuals. An assumption often violated in time series data due to inherent temporal dependencies. The parametric ESS estimators derived here can be seamlessly adapted to assess the significance of regression coefficients in linear regression models applied to time series data. Thus, our approach not only fortifies the foundation of correlation analysis in time series but also extends its robustness to the broader landscape of linear regression.

    \section*{Acknowledgments}
        The Wellcome Centre for Human Neuroimaging is supported by core funding from Wellcome [203147/Z/16/Z]. This research was partly supported by a grant from the French Minist\`ere de l’\'Education Nationale, de l’Enseignement Sup\'erieur et de la Recherche. The code related to this work is made openly available under an MIT license at \href{https://github.com/johmedr/corrts}{https://github.com/johmedr/corrts}.     
    
\bibliographystyle{rss}

\bibliography{references}
\appendix
    \section{Link with Rice's formula} 
        \label{sec:ess-rice}
        The second spectral moment is related to the average number of zero-crossings $N_{0}$ of the process by Rice's formula \cite{rice1944mathematical} (for derivation, see~\cite[Ch. 7.4, Example 7.9]{cox2017theory}):
         \begin{align}  
            N_{0}= n \sqrt{|\rho''(0)|}/\pi 
        \end{align} 
        This expression can be used to express the ESS in terms of expected number of zero-crossings. We observe that  
        \begin{align}
            \nu_\infty = \sqrt{\pi} N_0
        \end{align} 
        This gives a simple way to estimate the ESS of empirical time series from the mean number of zero-crossings.

    \section{Analytical ESS for Wavelet-based correlations}
        \label{sec:ess-wavelet}
        This section concerns testing correlations in band-limited power using time-frequency representations, particularly with Morlet wavelets, which align well with the ESS theory for quasi-Gaussian autocorrelation processes.
        
        Morlet wavelets, defined as the product of a complex sinusoid with a Gaussian, provide a time-frequency representation of a time series through the wavelet transform. Specifically, a wavelet $ w $ is given by \cite{cohen2019better}:
        \begin{align}
            w = \exp(-j 2 \pi f t -t^2/ (2 \sigma^2))
        \end{align}
        where $t$ is time, $ f $ is frequency, and $ \sigma $ is tied to the number of cycles $ N_c $ as $\sigma = N_c/2 \pi f$.
        
        For signals with a rapidly decaying autocorrelation compared to the Gaussian kernel width in the wavelet, the time-frequency representation remains relatively smooth. Considering a time series derived from the log-power (or other normalisation function) of a wavelet transform at a given frequency, the autocorrelation $ \rho $ is approximately Gaussian with second-spectral moment 
        \begin{align}
            |\rho''(0)| \approx  \frac{\pi^2 f^2}{N_c^2}
        \end{align}
    
        To determine the significance of the correlation between band-limited power, the number of cycles in Morlet wavelets can be adjusted, providing a direct assessment of the ESS for the correlations. If two series, derived from the same number of cycles but different frequencies $ f_1 $ and $ f_2 $, are correlated, their ESS is:
        \begin{align}
            \nu_\infty = \frac{n}{N_c} \sqrt{\frac{\pi}{2}}\sqrt{f_1^2 + f_2^2} 
        \end{align}
        For $ f_1 = f_2 = f $, the ESS is simply $ \nu_\infty = \sqrt{\pi} n f/N_c $. Importantly, the chosen number of cycles determines the ESS of power correlations.

\end{document}